\pdfoutput=1
\documentclass{JINST}

\title{Radiation Tolerance of 65 nm CMOS Transistors}

\author{M. Krohn$^a$, B. Bentele$^a$, D.C. Christian$^b$\thanks{Corresponding author}, J.P. Cumalat$^a$, G. Deptuch$^b$, F. Fahim$^b$, J. Hoff$^b$, A. Shenai$^b$, S.R. Wagner$^a$\\
\llap{$^a$}Department of Physics, University of Colorado,\\
  Boulder, Colorado 80309-0390, U.S.A.\\
\llap{$^b$}Fermi National Accelerator Laboratory,\\
  Batavia, IL 60510, U.S.A.\\
  E-mail: \email{dcc@fnal.gov}}

\abstract{
We report on the effects of ionizing radiation on 65 nm CMOS transistors held at approximately $-20 \,^{\circ}$C during irradiation.  The pattern of damage observed after a total dose of 1 Grad is similar to damage reported in room temperature exposures, but we observe less damage than was observed at room temperature.}


\begin{document}

\section{Introduction}

The need for extremely radiation tolerant electronics is one of the major issues confronting high energy physics in the era of High Luminosity running at the CERN \cite{CERN} Large Hadron Collider (HL-LHC).
Tests by Bonacini, $\textit{et al.}$ \cite{Bonacini} at CERN, published in 2012, established 65 nm CMOS as the leading candidate technology for HL-LHC electronics.  Using an X-ray beam, Bonacini, $\textit{et al.}$ exposed 65 nm transistors to a total dose of 200 Mrad.  Their results showed, with one exception, relatively small changes in transistor parameters for normal layout standard gate oxide thickness (core) transistors.  The exception was a dramatic loss of maximum drain-source current in the narrowest PMOS transistors.  The CERN group concluded that 65 nm CMOS technology could be used for HL-LHC applications with no special design considerations, except that all core devices should have width greater than 360 nm.

The RD53 collaboration was formed in 2014 to further explore the feasibility of using 65 nm CMOS technology to design a pixel readout chip for use at the HL-LHC \cite{RD53}.  
 The group established a total ionizing dose tolerance goal of 1 Grad.  The measurements reported in this paper were done in the context of RD53.  Discussions late in 2013 within RD53 centered on the fact that the data presented in reference \cite{Bonacini}, and also subsequent data collected by the CERN group and by a group from CPPM \cite{CPPM}, contain evidence of significant room temperature annealing during the time between X-ray exposures.  Both CMS and ATLAS currently plan to operate their HL-LHC pixel vertex detectors at approximately  $-20\,^{\circ}\mathrm{C}$.   
This choice is because the silicon strip trackers will operate at $-20\,^{\circ}\mathrm{C}$ in order to limit leakage current in the silicon sensors, which would otherwise require much more cooling and therefore more mass in the tracking volume.  Concern was expressed that because of reduced annealing, 65 nm circuits might experience greater radiation damage than had been observed in room temperature exposures if the circuits were maintained at  $-20\,^{\circ}\mathrm{C}$ during irradiation.

We report the results of an irradiation of 65 nm transistors performed using the Gamma Irradiation Facility\cite{GIF} at Sandia National Laboratories \cite{Sandia}.  The devices under test were maintained at a temperature $\lesssim -20\,^{\circ}\mathrm{C}$ during irradiation.

\section{Apparatus and Technique}

\subsection{Test ASIC}
\label{ssec2_1}
A 65 nm CMOS Application Specific Integrated Circuit (ASIC) containing individual transistors connected to wire bond pads was designed at Fermilab and fabricated by the Taiwan Semiconductor Manufacturing Company (TSMC)\cite{TSMC}.\footnote{Our test chip was fabricated at TSMC fab 14; the devices tested earlier at CERN were fabricated at TSMC fab 12\cite{SB}.}  The test ASIC was part of a multi-project wafer submitted to TSMC through the Metal Oxide Semiconductor Implementation Service (MOSIS)\cite{MOSIS}.
The chip was divided into two parts, one part intended primarily for lifetime studies of devices operated at liquid argon temperature, and one part intended for radiation tolerance testing.
Transistors intended for radiation tolerance testing were laid out in groups of similar transistors (for instance, NMOS transistors with channel length $L = 60$ nm and width W from 120 nm to 1000 nm). 
Within a group, all transistors share a diode-protected gate pad, and an (unprotected) source/drain pad.  The other drain/source of every transistor is connected to its own (unprotected) wire bonding pad.  We tested PMOS and NMOS core (1.2 V) transistors, and NMOS I/O (2.5 V) transistors (with double thickness gate oxide).

\subsection{ASIC package, test equipment, and measurement procedures}
\label{ssec2_2}

The test ASICs were wire bonded into (64-pin) pin grid array (PGA) chip carriers so that they could be irradiated on simple printed circuit boards (PCBs) containing only sockets for the ASICs and connectors for bias voltages. Transistor characteristics were measured 
by mounting one chip carrier at a time on a test board
containing switches that allowed individual transistors to be measured independently.  The number of pads on the test ASICs was too large to allow all pads to be wire bonded in one package, given the chosen chip carrier, so three different packages with different wire bonding patterns were made.  One package had bonds only to devices intended for cold tests.  NMOS transistors were wire bonded in the second package, and PMOS transistors were wire bonded in the third package.  The devices intended for cold tests are all large transistors unlikely to be used in a pixel readout ASIC.  They have been excluded from this analysis.

\begin{figure}
\begin{center}
\includegraphics[width=0.9\textwidth]{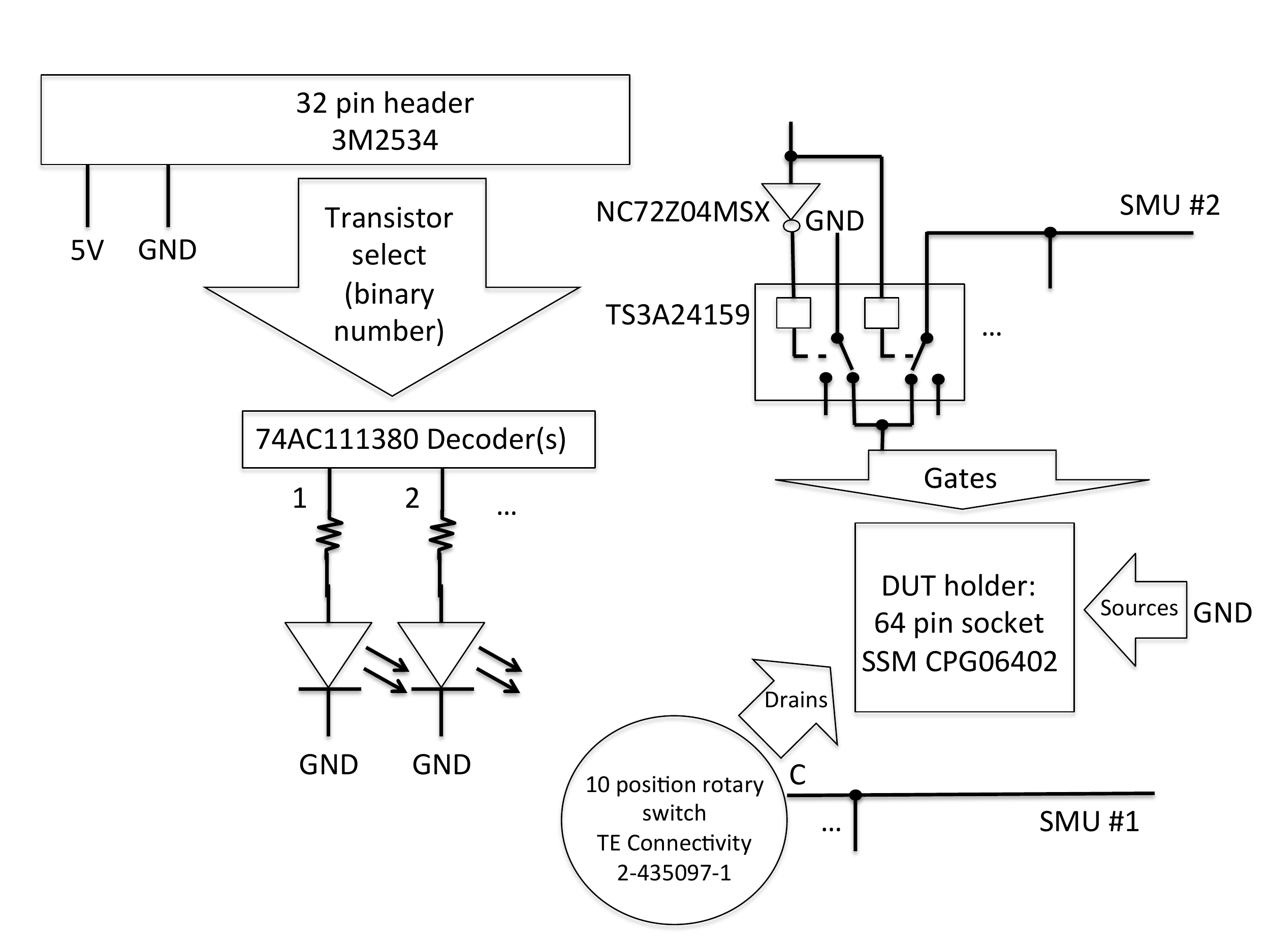}
\caption{Simplified schematic of the PCB used to measure NMOS transistor characteristics.}
\label{fig:testpcb}
\end{center}
\end{figure}

A different PCB was used to test each ASIC package. A simplified schematic of the PCB used to test NMOS transistors is shown in Figure~\ref{fig:testpcb}.  The PCB used to test PMOS transistors was very similar.
Two Keithley\cite{Keithley} 237 Source Measurement Units (SMUs)\cite{K237} were used, one to bias transistor gates, and one to measure drain-souce current.  A Labview\cite{Labview} program running on a laptop computer was used to sequence and control the measurements.  
The two SMUs were controlled via General Purpose Interface Bus (GPIB)\cite{GPIB}.
Logic on the test PCB was controlled via USB using a National Instruments\cite{NI} USB-6501 I/O board\cite{6501}, connected to the test PCB by a ribbon cable.
Bias voltage for the protection diodes was generated by a voltage regulator on the test PCB from the 5 V provided by the laptop USB port.  The Labview program controlled solid state switches on the test PCB that connected one of the SMUs to a single gate pad at a time; unused gates were grounded.  The program controlled LEDs on the test PCB to indicate how mechanical (rotary) switches on the test PCB should be set to connect the other SMU to a single transistor drain (unused drains were left floating).  All three voltage sources were referenced to a common ground plane on the test PCB, and the source pads for all transistors in a package were connected directly to this ground.  The fact that we did not separate the return current path for the two SMUs, together with possible parasitic circuits involving the protection diodes and the solid state switches in the OFF state, made it impossible for us to accurately measure the leakage current of transistors in the ASIC packages.

\subsection{Irradiation}
\label{ssec2_3}
\begin{figure}
\begin{center}
\includegraphics[width=0.6\textwidth]{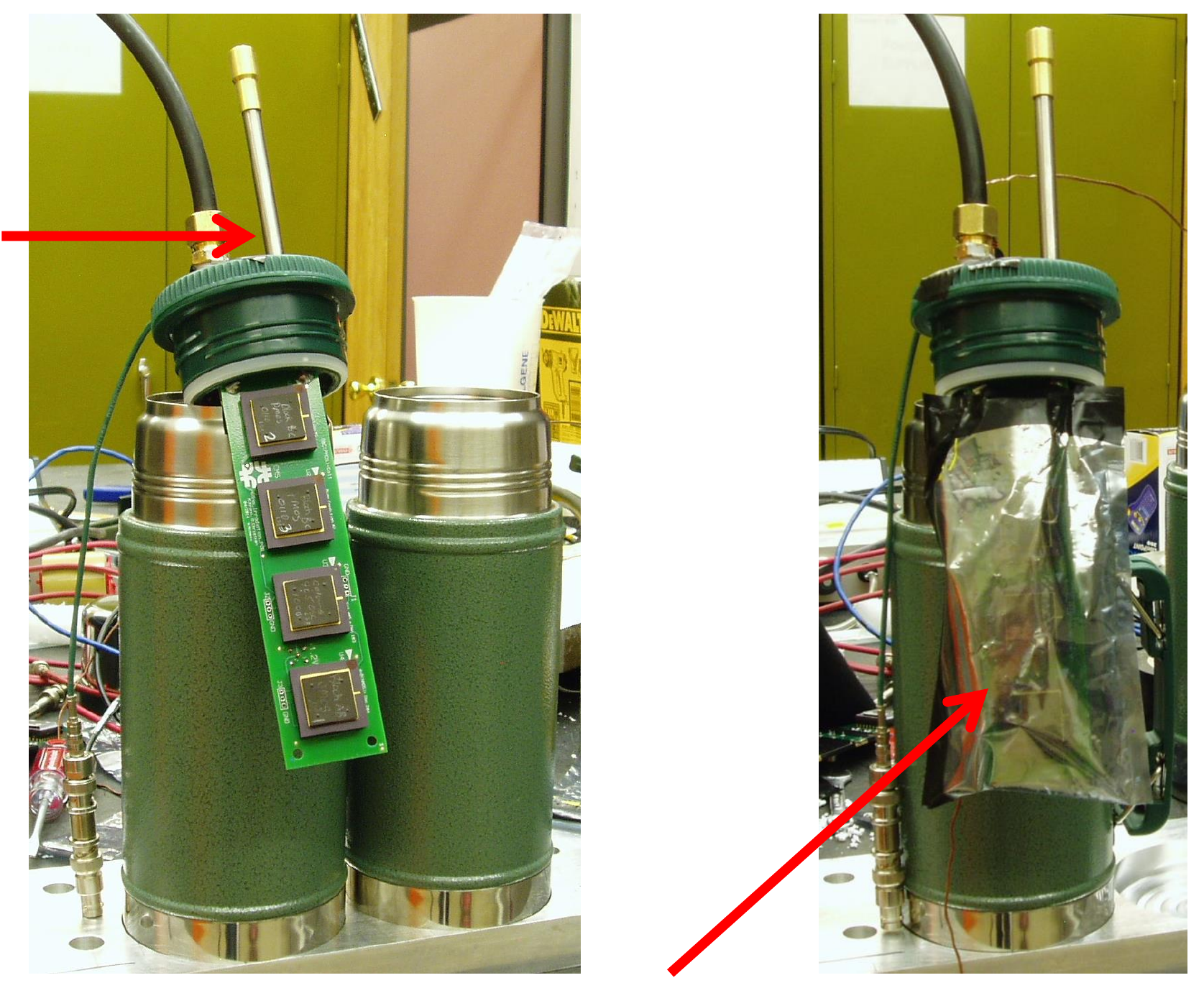}
\caption{Pictures are shown of a thermos bottle assembly, including an irradiation board with
four chip carriers, before insertion of the irradiation board into the thermos bottle.
In the left photo, the red arrow points to the vortex tube\cite{vortex} on top of the thermos bottle lid. In the right photo, the red arrow points to an antistatic bag which wraps
the irradiation board and (LEMO) low-voltage cable before irradiation. These
bags separate the boards and voltage cables from the not-very-dry thermos bottle 
environment, and provide protection from the metal thermos bottle wall (the test structures
are as close to the inner thermos bottle wall as is safe, but not touching).  During irradiation, copper pipe was used to deliver air to the vortex tubes.}
\label{fig:thermos}
\end{center}
\end{figure}

The Sandia National Laboratories Gamma Irradiation Facility (GIF) uses $^{60}$Co sources to provide controlled doses of ionizing radiation.  $^{60}$Co decays by beta decay to an excited state of $^{60}$Ni. $^{60}$Ni relaxes to the ground state by emitting two gamma rays of energy 1.17 and 1.33 MeV \cite{pdg}.  At the Sandia GIF, $^{60}$Co is held in stainless steel ``source pins'' that are $3/8$ inch diameter and 18 inches long.  A number of source pins are mounted in an array and to first order, none of the beta electrons escapes the steel source pins.  When not in use, the sources are kept at the bottom of an 18 foot deep pool of deionized water which provides shielding.  The facility has three shielded irradiation cells in a single high bay area above the shielding pool.  Each irradiation cell has an opening in the floor that allows a source array to be raised out of the water into the cell by an elevator.  The cell that was used in these irradiations contained an array of 40 source pins arranged in a straight line.  The array contained approximately 225 kCi of $^{60}$Co.  Our test ASICs were held inside stainless steel thermos bottles (see Figure~\ref{fig:thermos}) positioned approximately 2 inches from the face of the source array.\footnote{The standard practice for $^{60}$Co irradiation calls for the electrical devices being tested to be shielded with 1.5mm of lead followed by 0.7 - 1.0 mm of aluminum\cite{ASTM standard F1892} ``in order to minimize dose enhancement effects caused by low-energy scattered radiation.'' Our setup did not include a lead-aluminum shielding structure.}  Cooling was provided by vortex tube coolers \cite{vortex} mounted in holes drilled through the plastic thermos bottle lids.  

The dose rate was 1425 rad/second as measured by an ion chamber placed inside one of the thermos bottles.\footnote{All dosimetry was provided by Sandia National Laboratories.}  The uniformity of the radiation field was checked by irradiating thermoluminescent dosimeters (TLDs) taped to each of the chip carriers on the irradiation PCBs.  The TLDs were read at the Radiation Metrology Laboratory at Sandia National Laboratories.  The nonuniformity was measured to be less than 6\% RMS by comparing the truncated mean (middle two of four) of the four TLDs at each chip carrier position to the average of truncated means, for measurements taken at the start and end of the irradiation.
This variation, which we did not correct for because it showed no obvious pattern at the different chip carrier positions, dominates the error on the ion chamber measurement.
The TLD measurements also provided a check of the dose rate measured with the ion chamber.

During irradiation, gamma rays interacted in the walls of the thermos bottles and directly heated the inside of the thermos bottles.  In order to maintain the temperature of the test devices at less than $-20\,^{\circ}$C during long irradiations, especially during daytime when the outside temperature was $\sim35\,^{\circ}$C, it was necessary to precool the compressed air input to the vortex tubes and to insulate the copper tubes carrying air to the vortex tubes.  Figure 3 shows the temperature of the two thermos bottles during long irradiations.
Temperatures were measured using a K-type thermocouple in each thermos bottle, read out and recorded with a Fluke 52 II digital thermometer\cite{Fluke52}.  The calibration error for K-type thermocouples used near $-20\,^{\circ}$C is $\pm 2.2\,^{\circ}$C\cite{typeK}.
The precooling of the compressed air was improved after the first two long irradiations, during which the temperature in one of the two thermos bottles reached $-15\,^{\circ}$C.

\begin{figure}
\begin{center}
\includegraphics[width=5in]{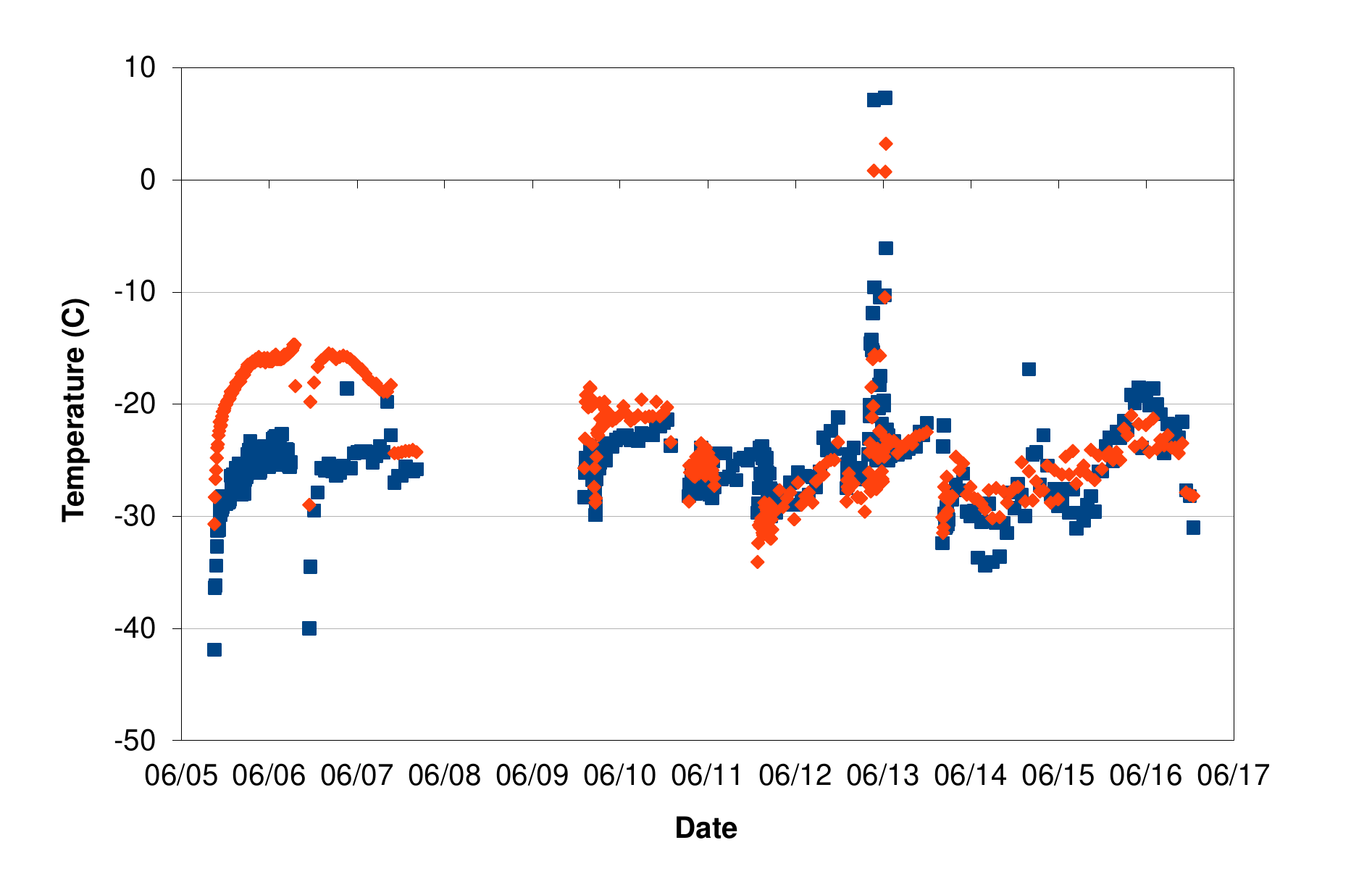}
\caption{The temperature measured inside the two thermos bottles ($\#1$ in blue and $\#2$ in red) during long irradiations.
No irradiation was performed during the day on (Saturday) June 8, or on June 9.
The two spikes where the temperature reached about $8\,^{\circ}$C in both thermos bottles for 30 minutes late on June 12 occurred because the compressed air unexpectedly shut off.}
\label{fig:Temperature_History}
\end{center}
\end{figure}

During irradiation the chip carriers were mounted in sockets on irradiation PCBs.  Each irradiation PCB held four chip carriers (see Figure~\ref{fig:thermos}), two for PMOS packages, and one each for NMOS and cold transistor packages.  Transistor bias voltages were provided by Keithley 237 SMUs (located outside the shielded irradiation cell) connected to the irradiation PCBs by 20 foot long triax cables. The PMOS transistors were biased in two different ways.  In one package, the drains, sources, and gates were held at 1.2 V and the substrate was grounded; the other package was biased with all the gates and the substrate grounded, while the drains and sources were held at 1.2 V.  The gates of both the core NMOS and the I/O NMOS were biased at 1.2 V; all other nodes were grounded.  Twelve irradiations were performed over 15 days, as shown in Table~\ref{tab:Irradiation_Schedule}.  After each irradiation step, a single characteristic curve was recorded for each transistor.  All measurements were made at room temperature. The drain-source voltage was set to 1.2 V and the drain-source current was measured
as the gate-source voltage was swept from 0 to 1.2 V.  It took $\sim$10 minutes to test the transistors in each package.  The ASIC packages were kept at $-20\,^{\circ}$C in a freezer when not being tested or irradiated.

\begin{table}
\begin{center}
\caption{The irradiation schedule, showing the 2 weeks it took to accumulate 1 Grad.}
\begin{tabular}{| p{2cm} | p{2.5cm} | p{2cm} | p{4cm} |}
\hline
Date & Length & Dose(Mrad) & Cumulative Dose(Mrad)\\ \hline
June 2 & 1 hour & 5 & 5\\ \hline
June 3 & 1 hour & 5 & 10 \\ \hline
June 3 & 1 hour 45 mins & 9 & 19 \\ \hline
June 3 & 4 hour 15 mins & 22 & 41 \\ \hline
June 4-5 & 12 hours & 62 & 103 \\ \hline
June 5-6 & 22 hours & 113 & 215 \\ \hline
June 6-7 & 22 hours & 113 & 329 \\ \hline
June 9-10 & 22 hours & 113 & 441 \\ \hline
June 10-11 & 17 hours & 87 & 528 \\ \hline
June 11-12 & 22 hours & 113 & 641 \\ \hline
June 12-13 & 22 hours & 113 & 754 \\ \hline
June 13-16 & 66 hours & 339 & 1093 \\
\hline
\end{tabular}
\label{tab:Irradiation_Schedule}
\end{center}
\end{table}

Pre-irradiation measurements of the transistors showed that a small number of transistors were broken either in fabrication or in the wire bonding process.  Approximately half of the transistors that were irradiated failed during the 15 days at Sandia.  One group of 12 NMOS transistors was broken mechanically by mishandling.
Most of the other transistors that failed also did so in groups, but without an obvious cause.  We replaced the package containing the group of 12 failed NMOS transistors
partway through the irradiation.  The replacement package received a total dose of 878 Mrad.
Tables~\ref{tab:nmosbroken} and~\ref{tab:pmosbroken} list all of the transistors included in this study and note which transistors failed and when the failures occurred.
Broken transistors were easily identified.  For many, the drain-source current was either very small or very large, independent of gate bias.  For a smaller number, the drain-source current varied approximately linearly with gate bias.


The most likely cause of transistor failures is electrostatic discharge (ESD).  We took a number of steps to reduce the probability of ESD, but our procedures had some deficiencies.  The chip carrier packages were transported in an antistatic box and when a package was mounted on, or removed from a PCB, the work was done on a grounded antistatic mat by a person wearing a wrist grounding strap.  The PCB was grounded before a chip carrier was inserted into or removed from a socket, but no ESD precautions were taken when the irradiation PCBs were inserted into the thermos bottles or when bias cables were connected.  All transistor gate pads were diode protected on-chip, but none of the source or drain pads was ESD protected.  Moreover, bias for the on-chip protection diodes was provided through only one pin of the PGA chip carriers.  If this pin failed to make contact before other pins while a package was being inserted into a socket, the protection diodes may not have been biased when they were needed most.

\begin{table}
\begin{center}
\caption{NMOS transistors: Each entry in one of the last three columns corresponds to a transistor and indicates the dose accumulated before the transistor was broken.  Transistors that were not broken have no entry.  Zero indicates a transistor that was broken before irradiation.  Transistors in the upper part of the table have standard thickness gate oxide; those in the lower part have gate oxide that is twice normal thickness.  All transistors are standard layout unless otherwise indicated; ELT indicates enclosed layout.  Transistors that share a gate pad are grouped together.  IC2 received the full dose of 1.1 Grad; IC3 was replaced by IC1 after a dose of 215 Mrad, so IC1 received 878 Mrad.  IC1 and IC3 were irradiated in thermos $\#1$; IC2 was irradiated in thermos $\#2$.}
\begin{tabular}{| p{2cm} | p{3cm} | p{1cm} | p{2cm} | p{2cm} | p{2cm} |}
\hline
W/L (nm) & Type (if not simple) & Gate & IC1 & IC2 & IC3 \\ \hline
\hline
120/60 & & 1 & 0 & 328 & 43 \\ \hline
240/60 & & 1 & & 5 & 43 \\ \hline
360/60 & & 1 & & 5 & 43 \\ \hline
480/60 & & 1 & & 5 & 0 \\ \hline
600/60 & & 1 & 0 & 5 & 43 \\ \hline
1000/60 & & 1 & & 754 &  \\ \hline
5000/500 & & 2 & & 754 &  \\ \hline
5000/5000 & & 2 & 426 & 754 &  \\ \hline
120/60 & Triple well & 2 & & 328 &  43 \\ \hline
5000/60 & Triple well & 2 & & 328 &  43 \\ \hline
1500/300 & Zero $V_t$ & 2 & & 5 &  43 \\ \hline
2050/60 & ELT & 2 & 0 & 5 &  43 \\ \hline
2240/300 & Zero $V_t$ ELT & 2 & & 328 &  43 \\ \hline
\hline
400/280 & & 3 & & 754 &  \\ \hline
500/280 & & 3 & & 754 &  \\ \hline
800/280 & & 3 & & 754 &  \\ \hline
1000/280 & & 3 & & 754 &  \\ \hline
5000/500 & & 3 & & 754 &  \\ \hline
5000/5000 & & 3 & & 754 &  \\ \hline
2220/280 & ELT & 3 & & 754 &  \\ \hline
3380/1200 & Zero $V_t$ ELT & 3 & & 754 &  \\ \hline
400/280 & Triple well & 3 & 426 & 754 &  \\ \hline
800/280 & Triple well & 3 & & 754 &  \\ \hline
\end{tabular}
\label{tab:nmosbroken}
\end{center}
\end{table}

\begin{table}
\begin{center}
\caption{PMOS transistors: Each entry in one of the last four columns corresponds to a transistor and indicates the dose accumulated before the transistor was broken. Transistors that were not broken have no entry.
IC4 and IC6 were biased with $V_s=V_d=V_g$. IC5 and IC7 were biased with $V_s=V_d=1.2$ V and $V_g=$GND.
All four packages received the full dose of 1.1 Grad.  IC4 and IC5 were irradiated in thermos $\#2$; IC6 and IC7 were irradiated in thermos $\#1$.}
\begin{tabular}{| p{2cm} | p{1cm} | p{2cm} | p{2cm} | p{2cm} | p{2cm} |}
\hline
W/L (nm) & Gate & IC4 & IC5 & IC6 & IC7 \\ \hline
\hline
120/60 & 1 & 0 & & 218 & 531\\ \hline
360/60 & 1 & 328 & & 13 & \\ \hline
600/60 & 1 & 0 & 754 & 43 & 443 \\ \hline
1000/60 & 1 & 5 & & 531 & \\ \hline
5000/500 & 2 & 0 & & & \\ \hline
5000/5000 & 2 & 328 & & 531 & \\ \hline
\end{tabular}
\label{tab:pmosbroken}
\end{center}
\end{table}

After the irradiations, the devices were kept at $-20\,^{\circ}$C in a freezer that could be powered either by 120 V or by 12 V and transported to Fermilab. 
Once at Fermilab the transistors were removed from the freezer and kept at room temperature for one week.   Multiple measurements were taken during this time. Then the transistors were held in an oven at $100\,^{\circ}$C for another week and a final set of measurements was made. This annealing schedule can be seen in Table~\ref{tab:Annealing_Schedule}. The transistors were not biased during transport or annealing.

\begin{table}
\begin{center}
\caption{The annealing times and temperatures of the transistors.}
\begin{tabular}{| p{3cm} | p{3cm} | p{3cm} | }
\hline
\multicolumn{3}{|c|}{Annealing Schedule} \\
\hline
June 16-24 & $-20^{\circ}$C & 8 Days\\ \hline
June 24 - July 1 & Room Temperature & 7 Days\\ \hline
July 1-8 & $100\,^{\circ}$C & 7 Days\\
\hline
\end{tabular}
\label{tab:Annealing_Schedule}
\end{center}
\end{table} 

\section{Analysis and Results}

\begin{figure}
\begin{center}
\includegraphics[height=7.12cm]{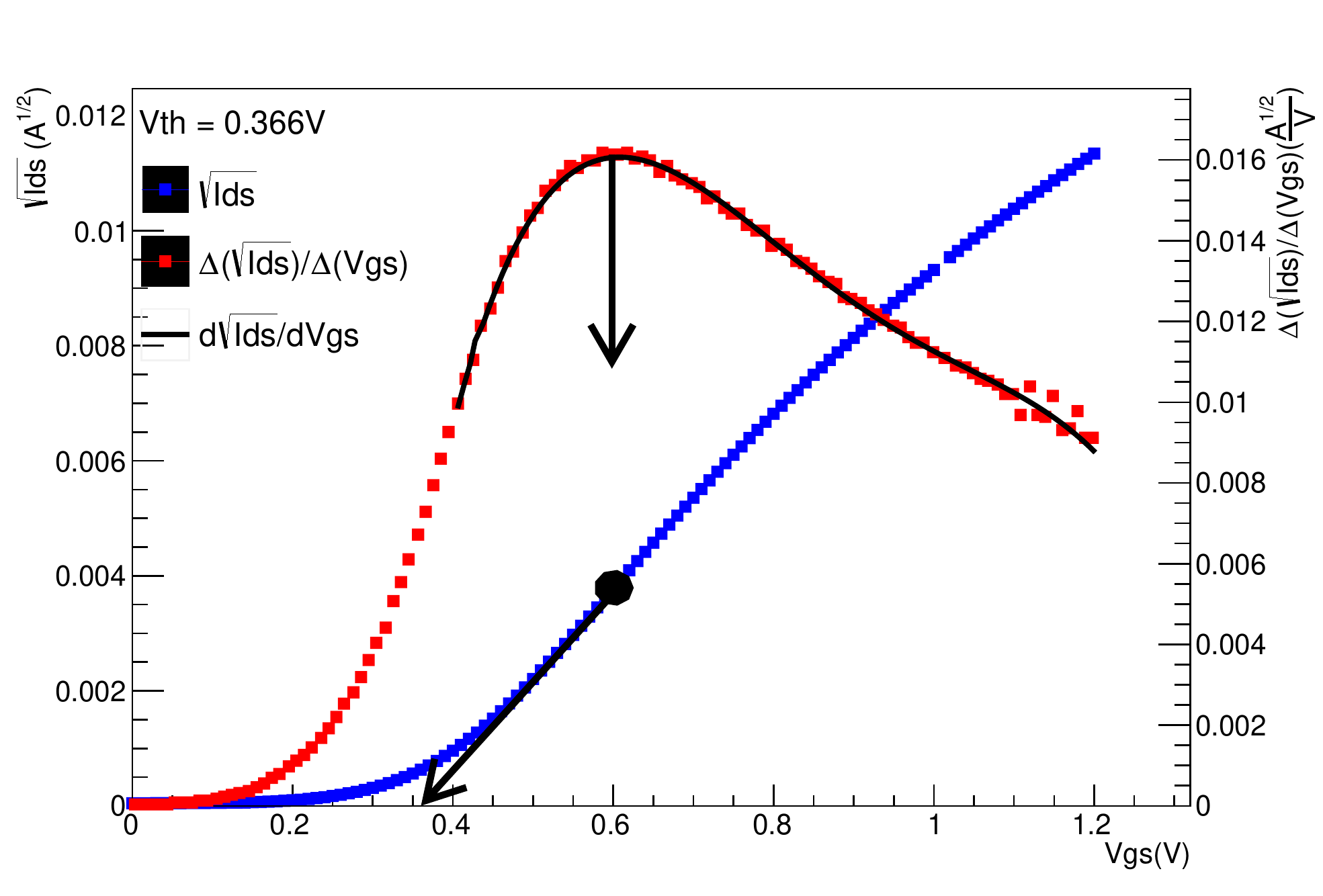}
\caption{This figure illustrates the quadratic extrapolation method used to determine the (saturation) threshold voltage ($V_{th}$) of an NMOS transistor.  The data shown is from the pre-irradiation measurement of the 240/60 transistor in IC3.
For PMOS transistors, $|I_{ds}|$ is used since $I_{ds}$ is negative.}
\label{fig:QuadMethod}
\end{center}
\end{figure}

Two quantities were extracted from each transistor characteristic: the maximum drain-source current and the (saturation) threshold voltage $V_{th}$.  The quadratic extrapolation method was used to determine the threshold voltage\cite{Schroder}.  As shown in Figure~\ref{fig:QuadMethod}, $V_{th}$ is defined to be the voltage at which a line tangent to the curve $\sqrt{|I_{ds}|}$ vs $ V_{gs}$ at the point of maximum $\frac{d\sqrt{|I_{ds}|}}{dV_{gs}}$ intercepts the $I_{ds}=0$ axis.  We determined the slope of the curve by fitting it with a fifth order polynomial and differentiating the fit function.  In Figure~\ref{fig:QuadMethod}, the red squares were computed using finite differences
\Big({${\sqrt{I_{ds}(N+1)}-\sqrt{I_{ds}(N)}\over{V_{gs}(N+1)-V_{gs}(N)}}$} \Big); the black line is the result of differentiating the fit to the curve $\sqrt{|I_{ds}|}$ vs $ V_{gs}$.

\begin{figure}
	\begin{minipage}[b]{.5\linewidth}
	\includegraphics[width=\linewidth]{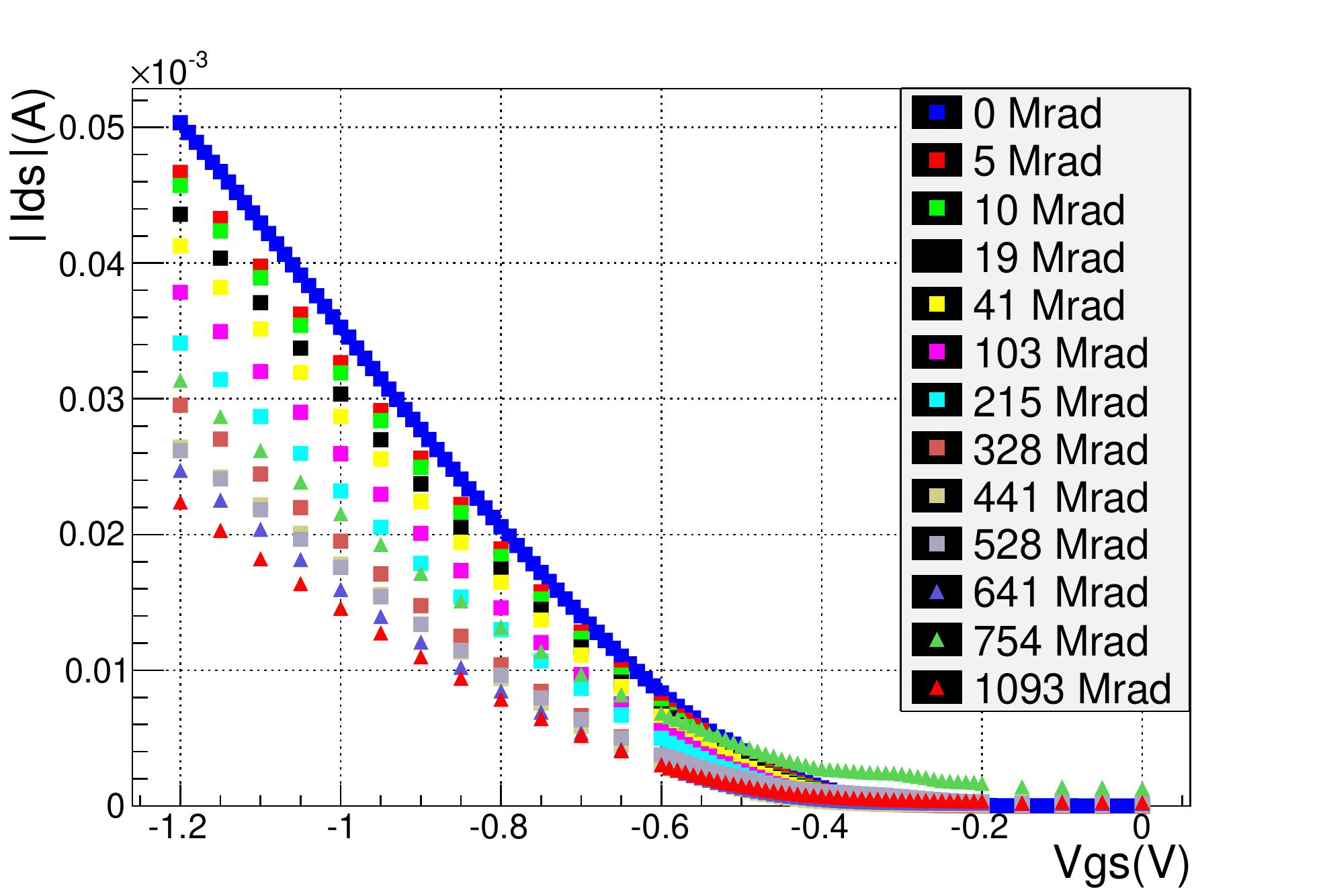}
	\end{minipage}
	\begin{minipage}[b]{.5\linewidth}
	\includegraphics[width=\linewidth]{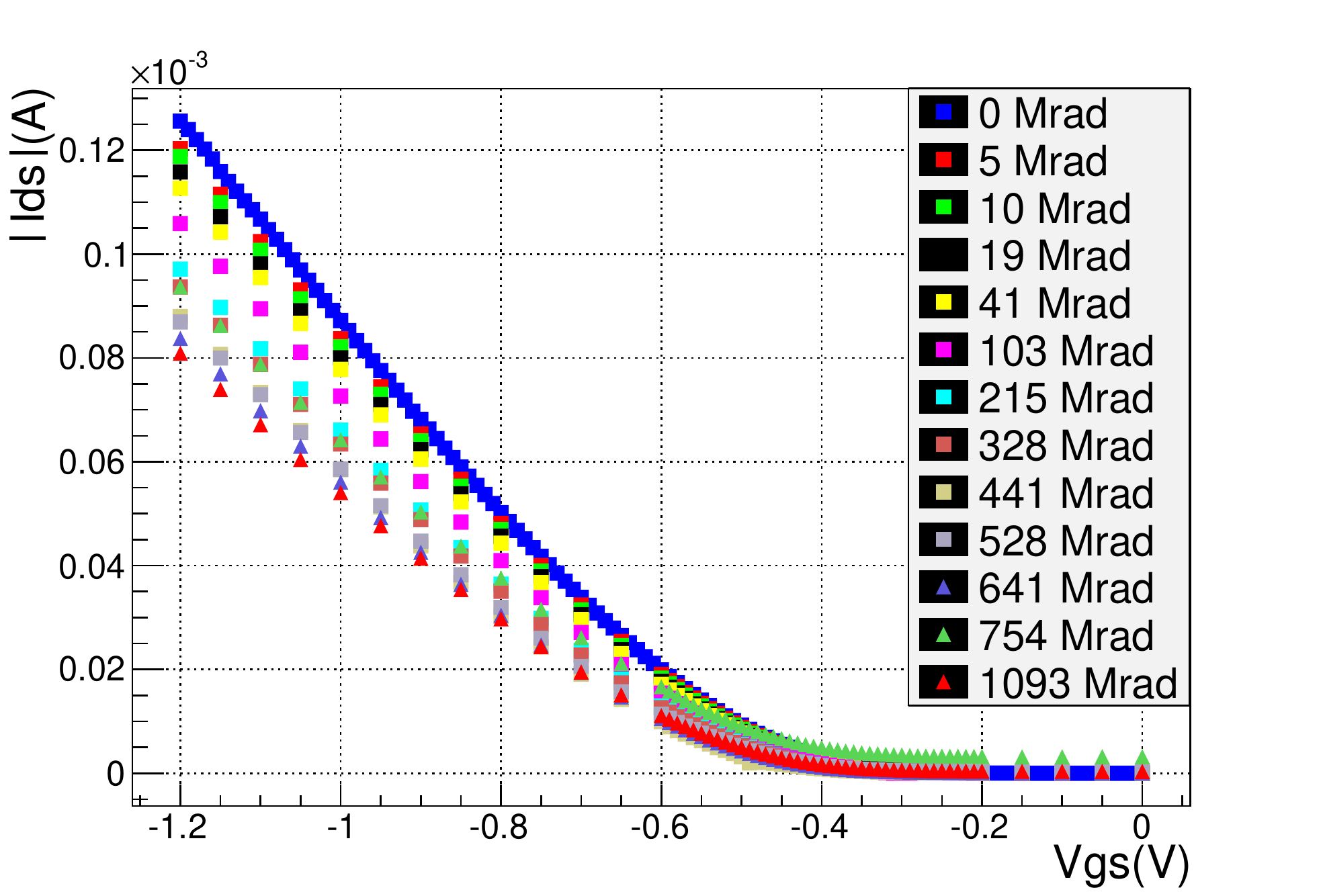}
	\end{minipage}
	\begin{minipage}[b]{.5\linewidth}
	\includegraphics[width=1\linewidth]{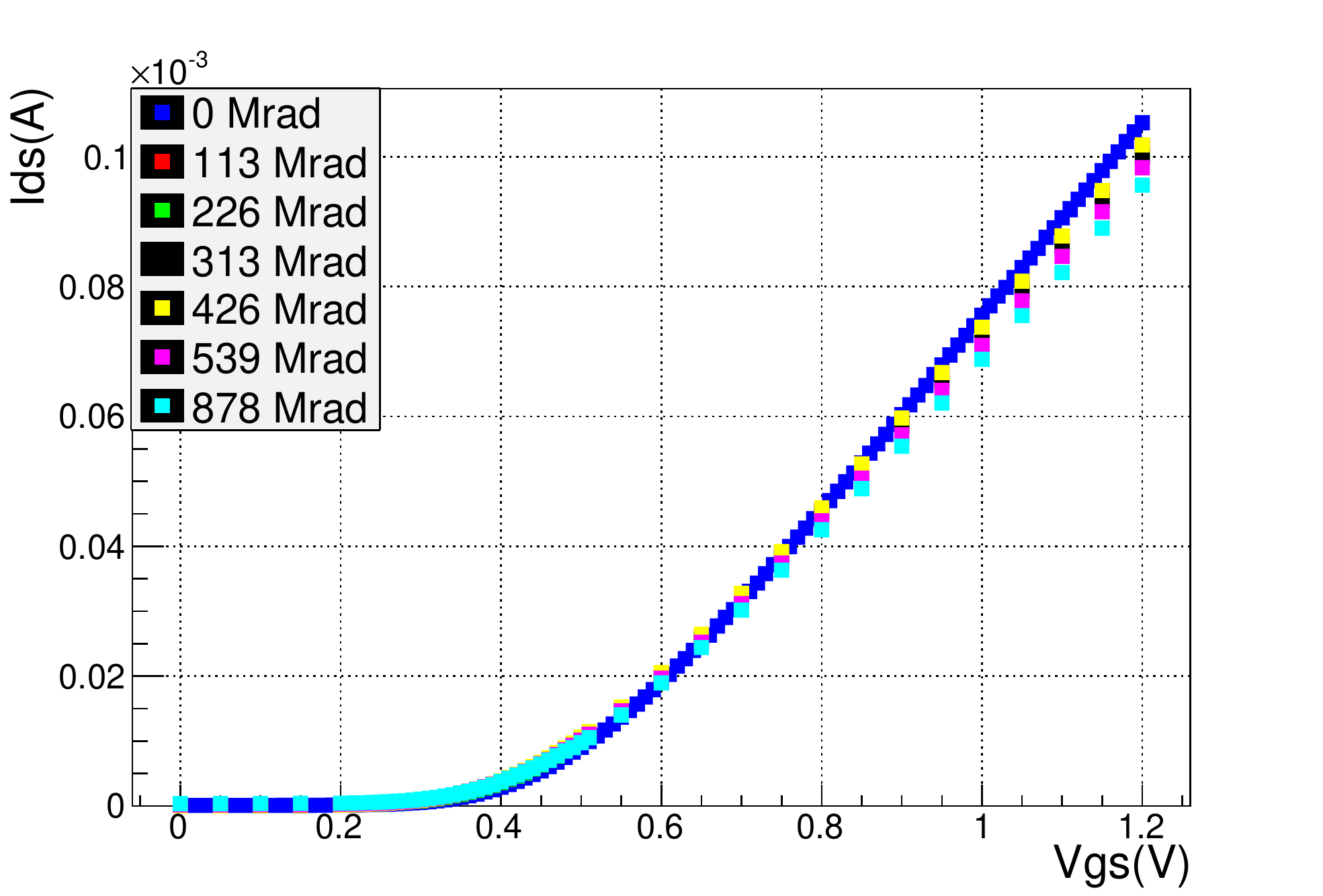}
	\end{minipage}
	\hfill
	\begin{minipage}[b]{.5\linewidth}
	\includegraphics[width=\linewidth]{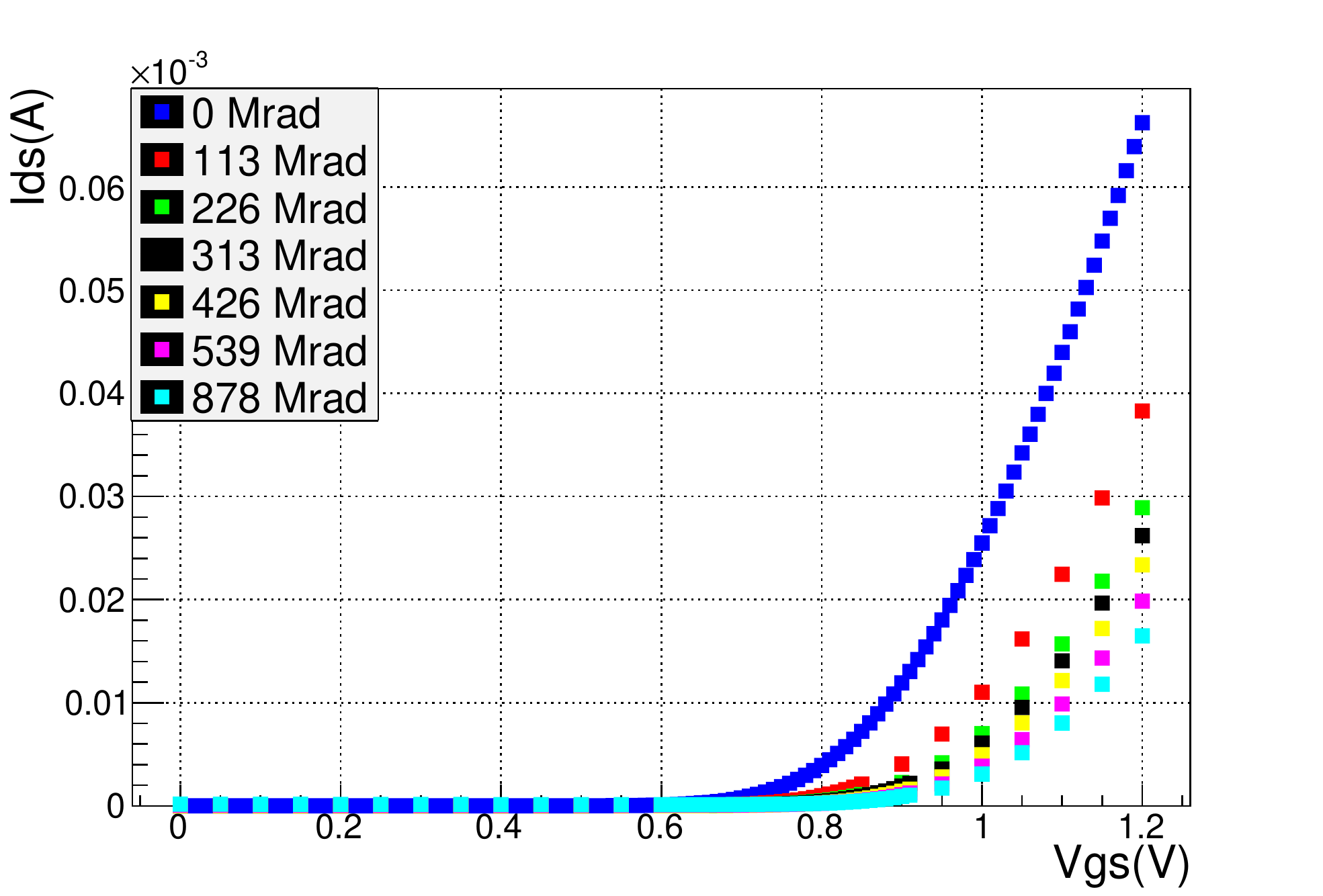}
	\end{minipage}
\caption{Transistor characteristic curves for total dose up to 1.1 Grad of (upper left) a 120/60 core PMOS, (upper right) a 360/60 core PMOS, and for total dose up to 878 Mrad of (lower left) a 240/60 core NMOS, and (lower right) a 1000/280 2.5 V NMOS.}
\label{fig:SuperpositionPlots}
\end{figure}

Figure~\ref{fig:SuperpositionPlots} illustrates the radiation effects observed in our data.  The most prominent effect is a decrease of the maximum drain-source current of core PMOS transistors.  The fractional decrease is largest for the smallest PMOS transistors; the maximum drain-source current of the smallest PMOS decreased by more than a factor of two.  The maximum drain-source current of core NMOS transistors also decreased, but only by $\sim5-10\%$.  No significant threshold shift was observed for any of the core transistors, but the threshold voltage of NMOS I/O transistors increased by 100 - 200 mV.  No error bars are included in the figures because the uncertainty in the SMU measurements is smaller than the symbols used to plot the measurements.

\begin{figure}
\begin{minipage}[b]{0.5\textwidth}
	\centering
	\includegraphics[width=\linewidth]{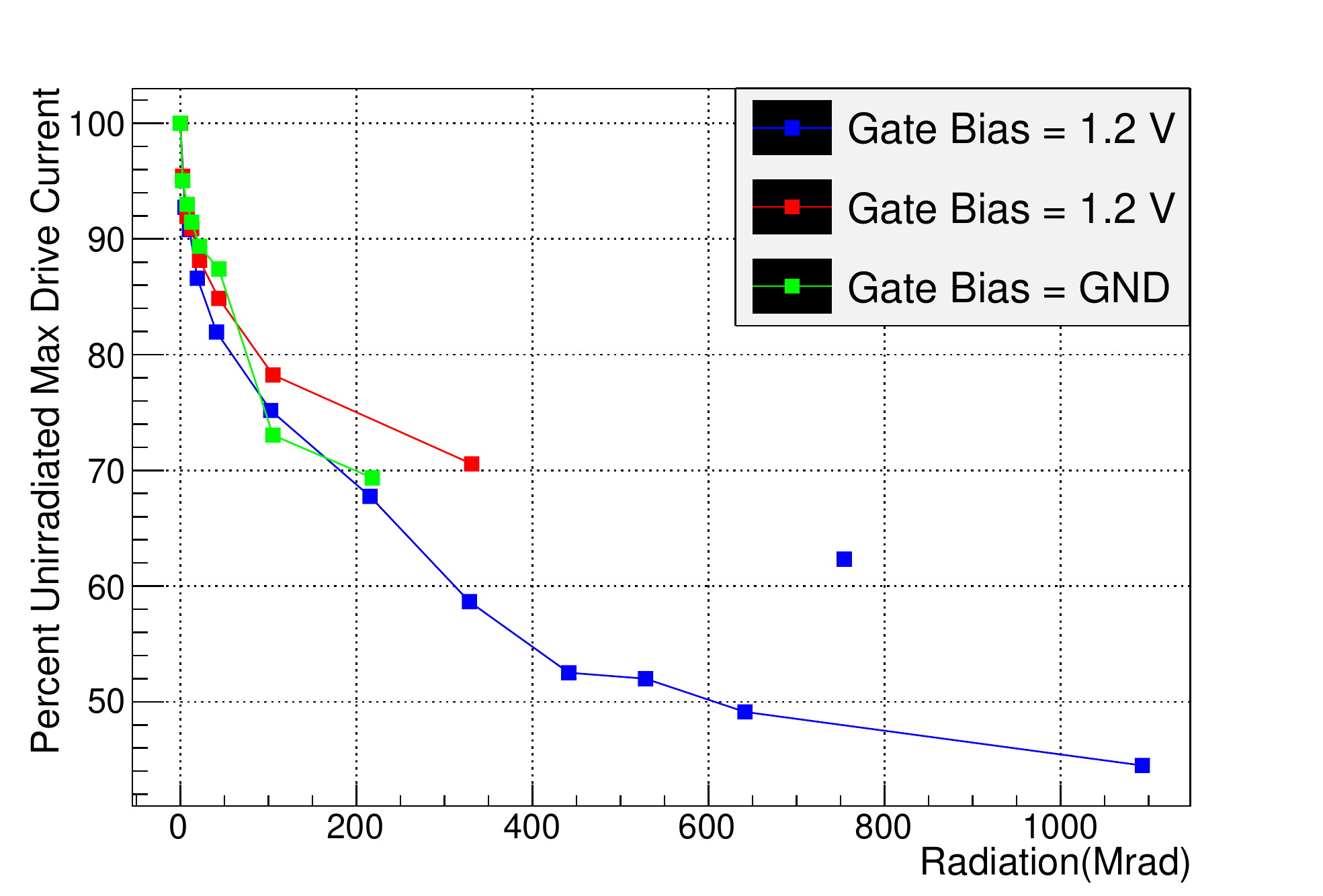}
\end{minipage}
\hspace{0.5cm}
\begin{minipage}[b]{0.5\textwidth}
	\centering
	\includegraphics[width=\linewidth]{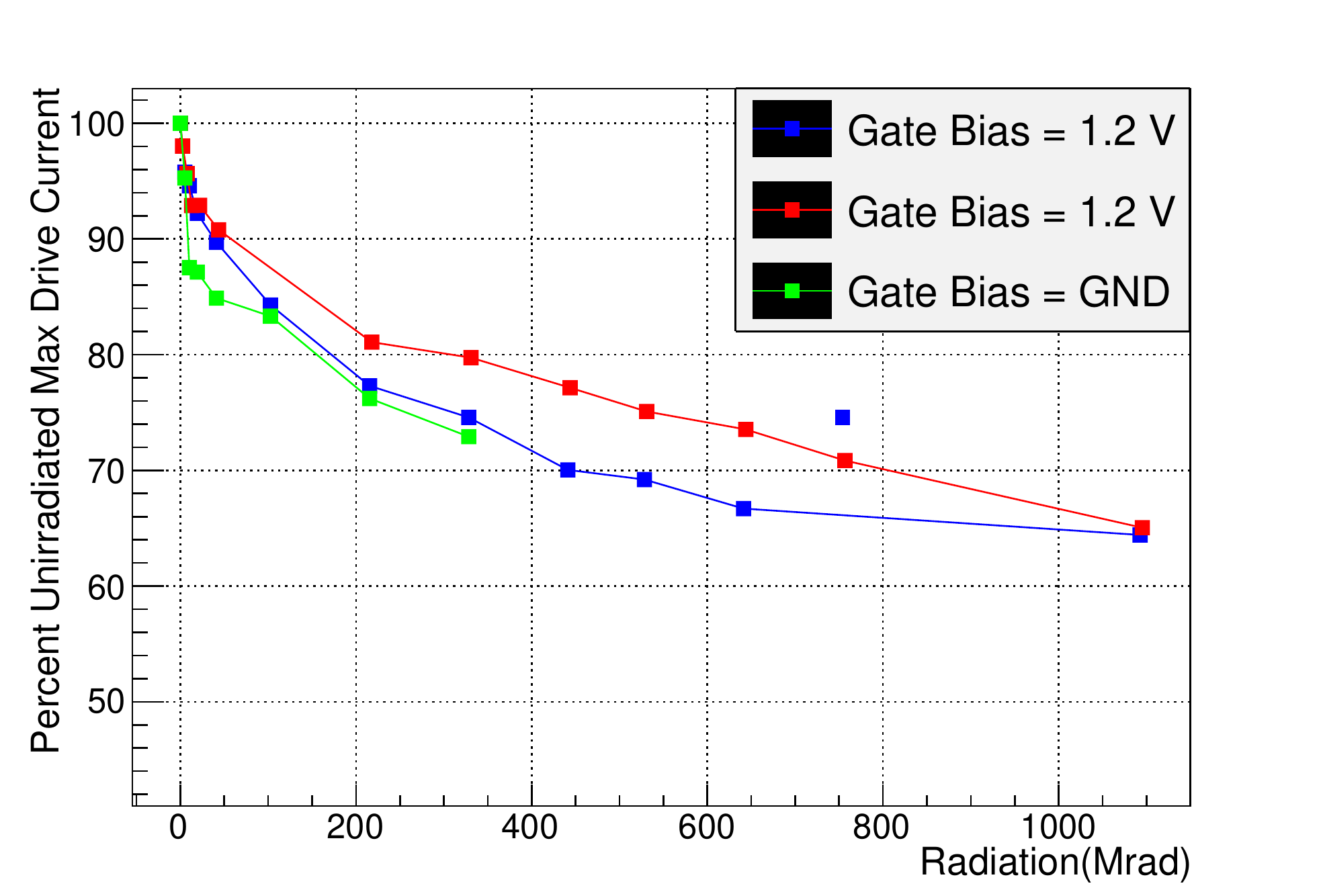}
\end{minipage}
\caption{The change in maximum drain-source current for similar PMOS core transistors irradiated with different gate bias voltages. The graph on the left is for 120/60 transistors and the graph on the right is for 360/60 transistors.  The lines connecting points do not represent a fit, and are included only to make the plots easier to read.  The transistor characteristics measured for transistors in package IC5 after 754 Mrad was accumulated were all offset by current not likely to have passed through the transistors (this can be seen in Figure 5).  Lines are not drawn through these points.  The most likely source of these offsets is leakage current due to moisture caused by condensation on the cold IC package.}
\label{fig:PMOSBiasConditions}
\end{figure}


\begin{figure}
\begin{minipage}[b]{0.5\textwidth}
	\centering
	\includegraphics[width=\linewidth]{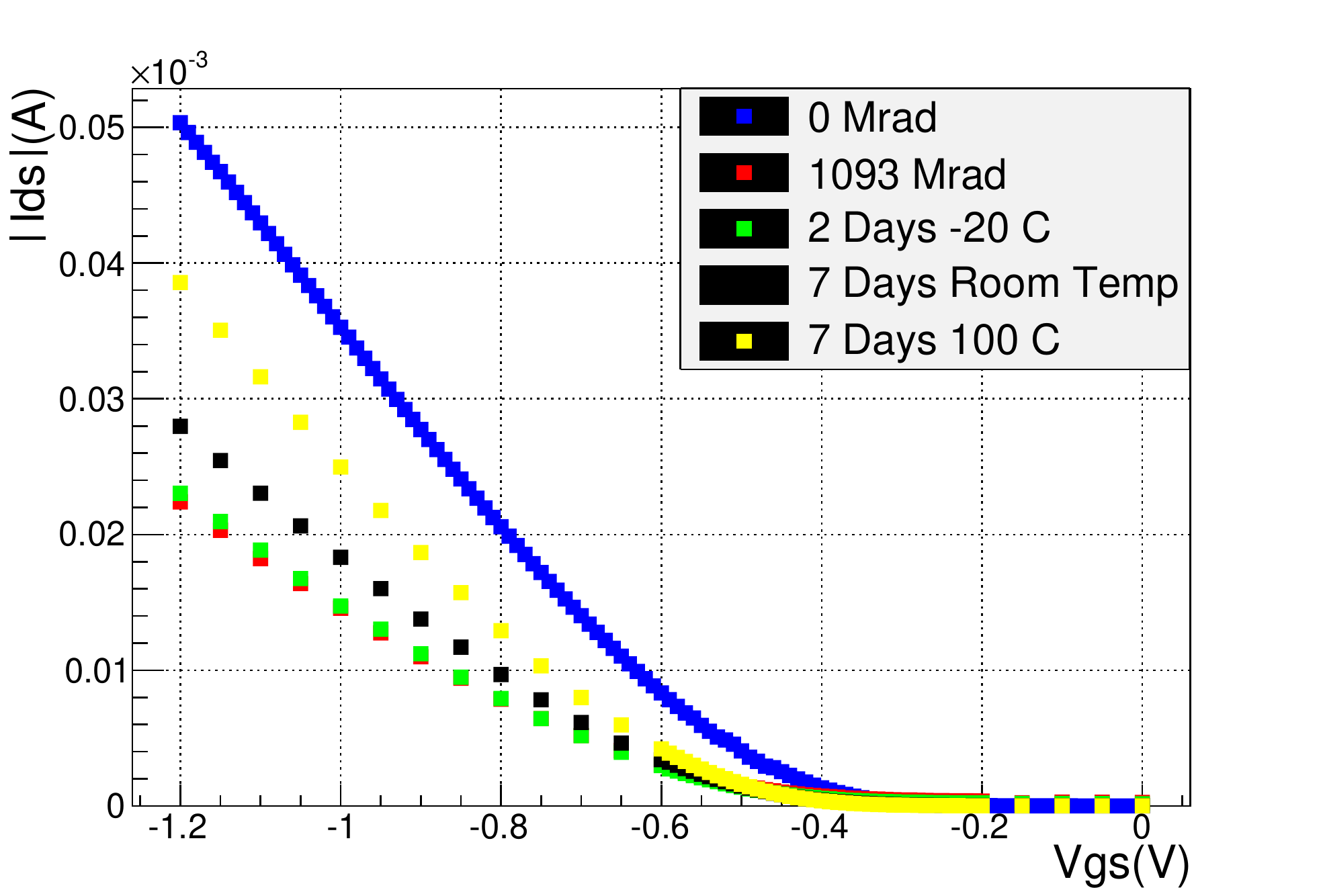}
\end{minipage}
\hspace{0.5cm}
\begin{minipage}[b]{0.5\textwidth}
	\centering
	\includegraphics[width=\linewidth]{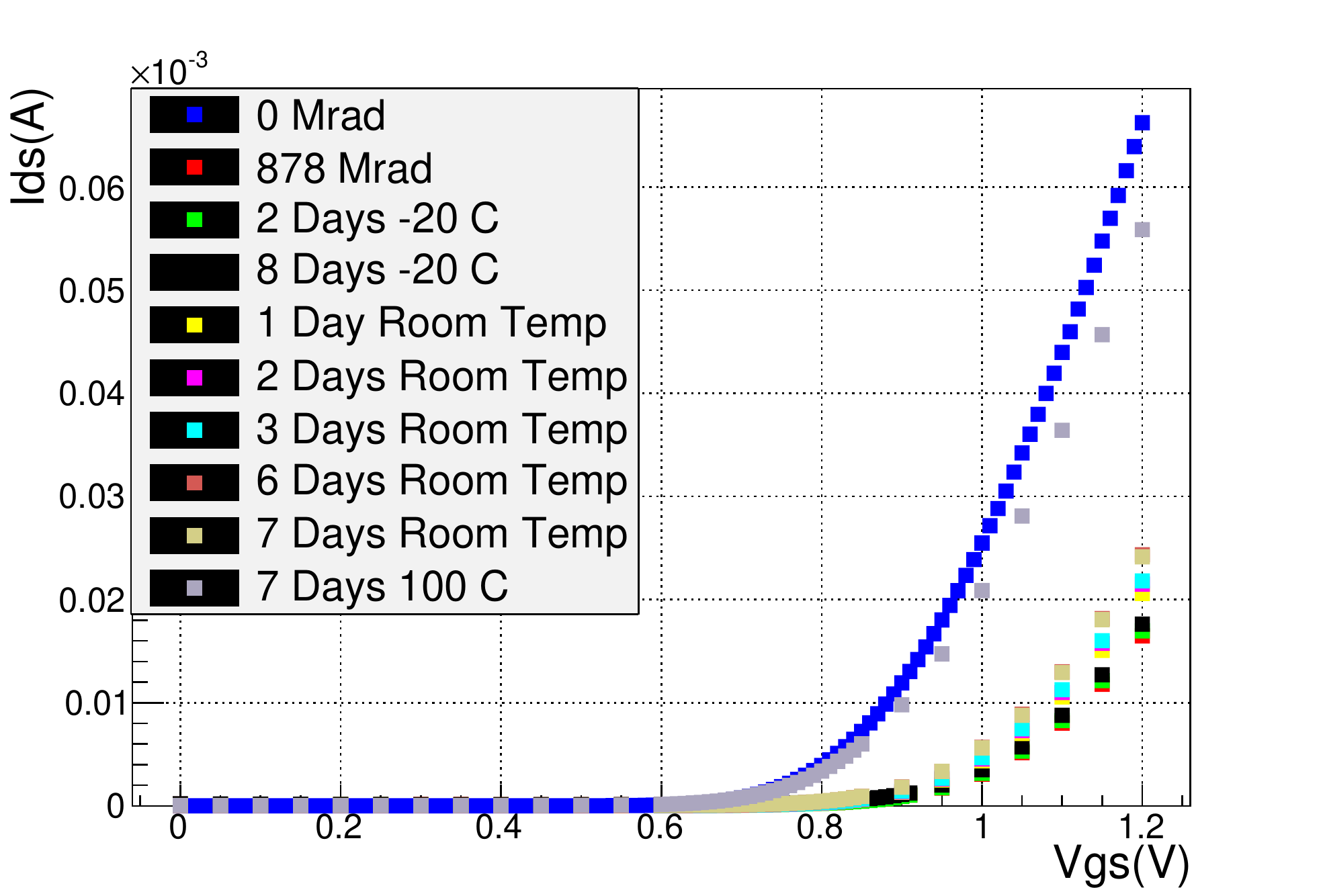}
\end{minipage}
\caption{Transistor chararcteristic curves during the annealing period for (left) a 120/60 core PMOS and (right) a 1000/280 2.5 V NMOS.}
\label{fig:AnnealSuperpositionPlots}
\end{figure}

No significant difference was observed between the radiation-induced changes of PMOS transistors biased during the irradiation with the gate in the ON state and PMOS transistors biased with the gate in the OFF state.  This is illustrated in Figure~\ref{fig:PMOSBiasConditions}.

\begin{figure}
\begin{minipage}[b]{0.5\textwidth}
	\centering
	\includegraphics[width=\linewidth]{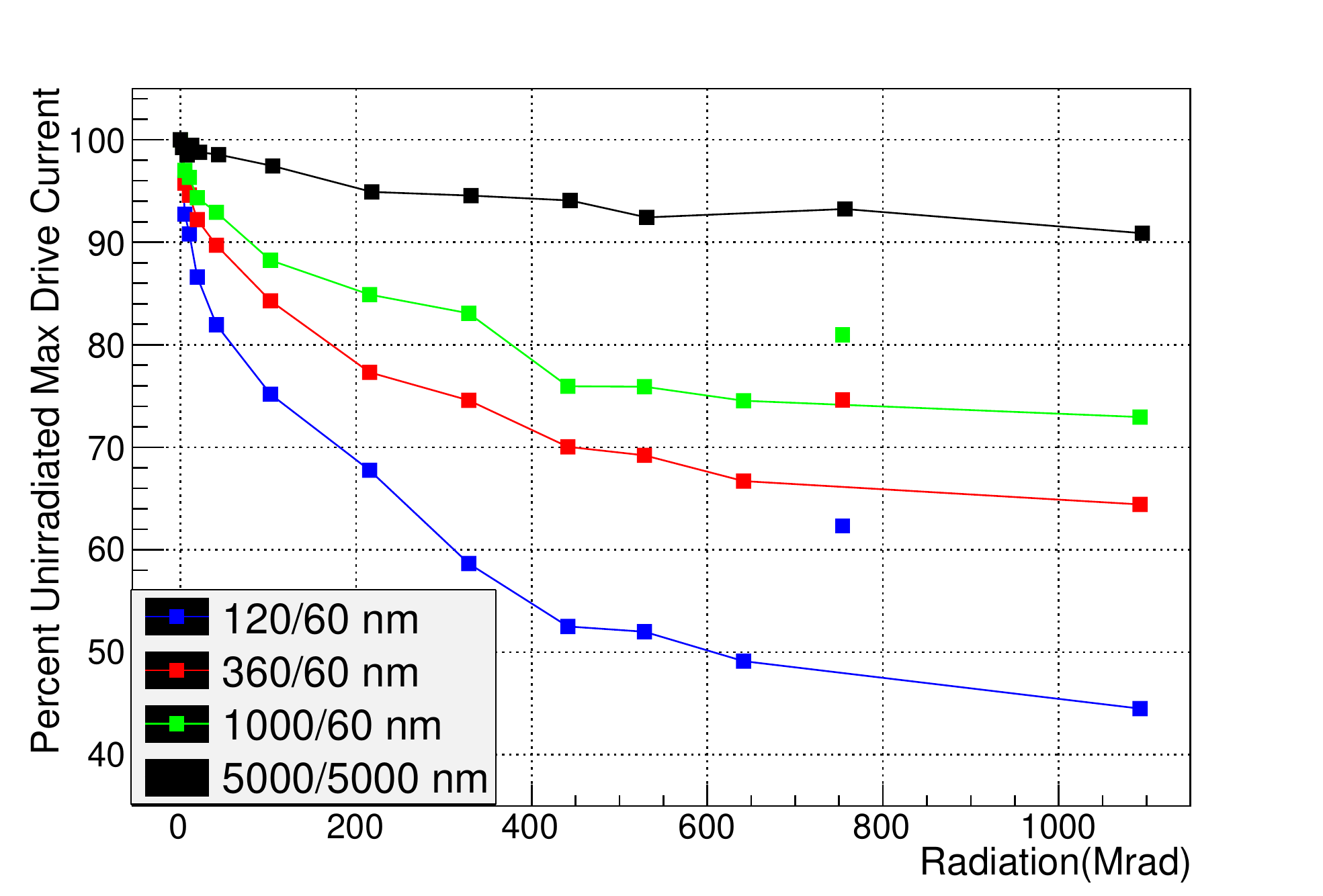}
\end{minipage}
\hspace{0.5cm}
\begin{minipage}[b]{0.5\textwidth}
	\centering
	\includegraphics[width=\linewidth]{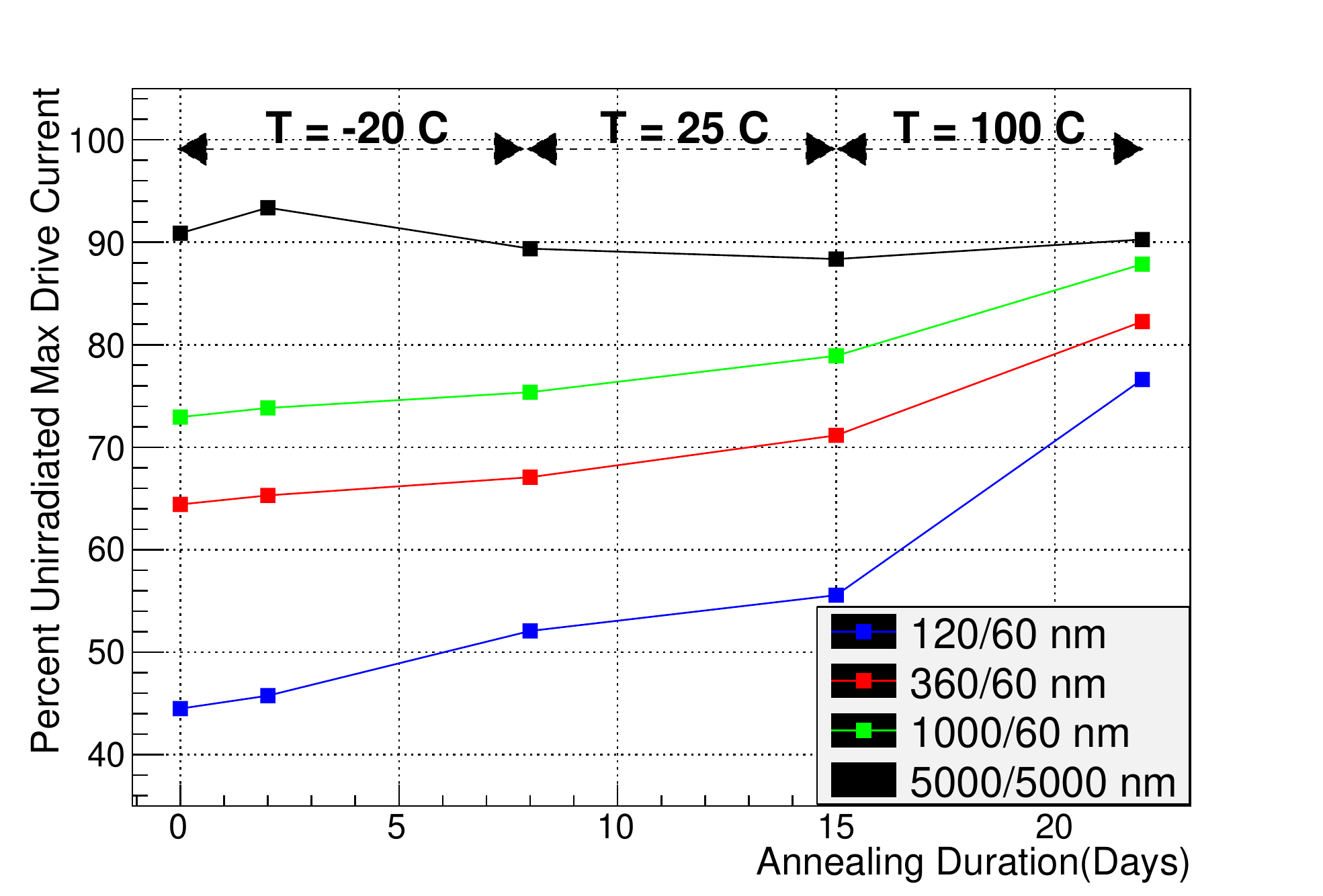}
\end{minipage}
\caption{The graph on the left shows the loss of maximum drain-source current during irradiation for 4 PMOS core transistors. The graph on the right shows the recovery of maximum drain-source current for the same 4 transistors during and after annealing.
As in Figure 6, lines are included to make the plots easier to read.
Once again, lines are not drawn through the points corresponding to measurements made after 754 Mrad of transistors in IC5.  The measurements shown for the 5000/5000 transistor are for the transistor in IC7.  For this transistor, no point is included corresponding to an integrated dose of 641 Mrad; we believe that the rotary switch was not set correctly during the measurement of this transistor characteristic since the recorded drain-source current was very small for all values of the gate bias.}
\label{fig:MaxCurDrive_PMOS}
\end{figure}

\begin{figure}
\begin{minipage}[b]{0.5\textwidth}
	\centering
	\includegraphics[width=\linewidth]{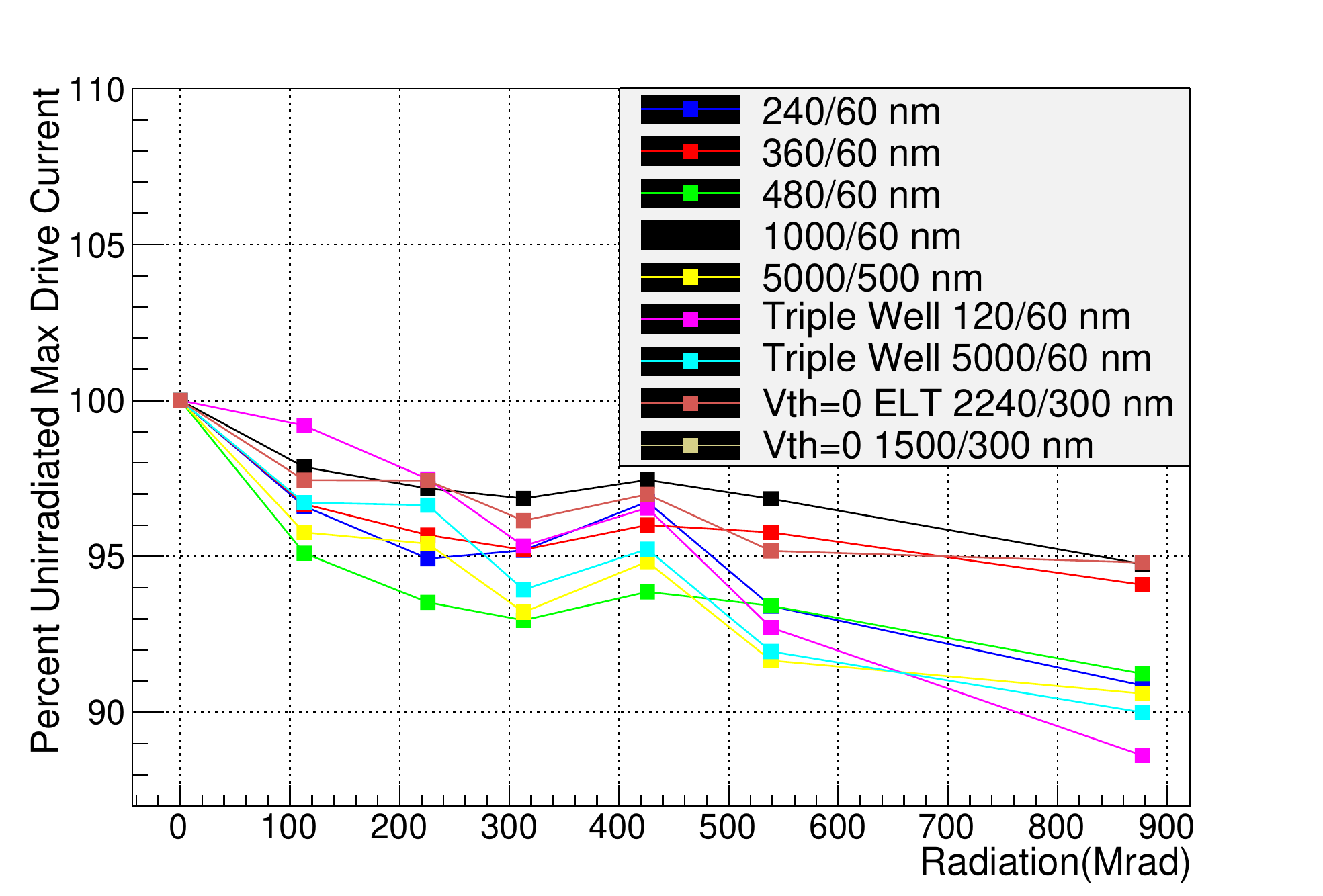}
\end{minipage}
\hspace{0.5cm}
\begin{minipage}[b]{0.5\textwidth}
	\centering
	\includegraphics[width=\linewidth]{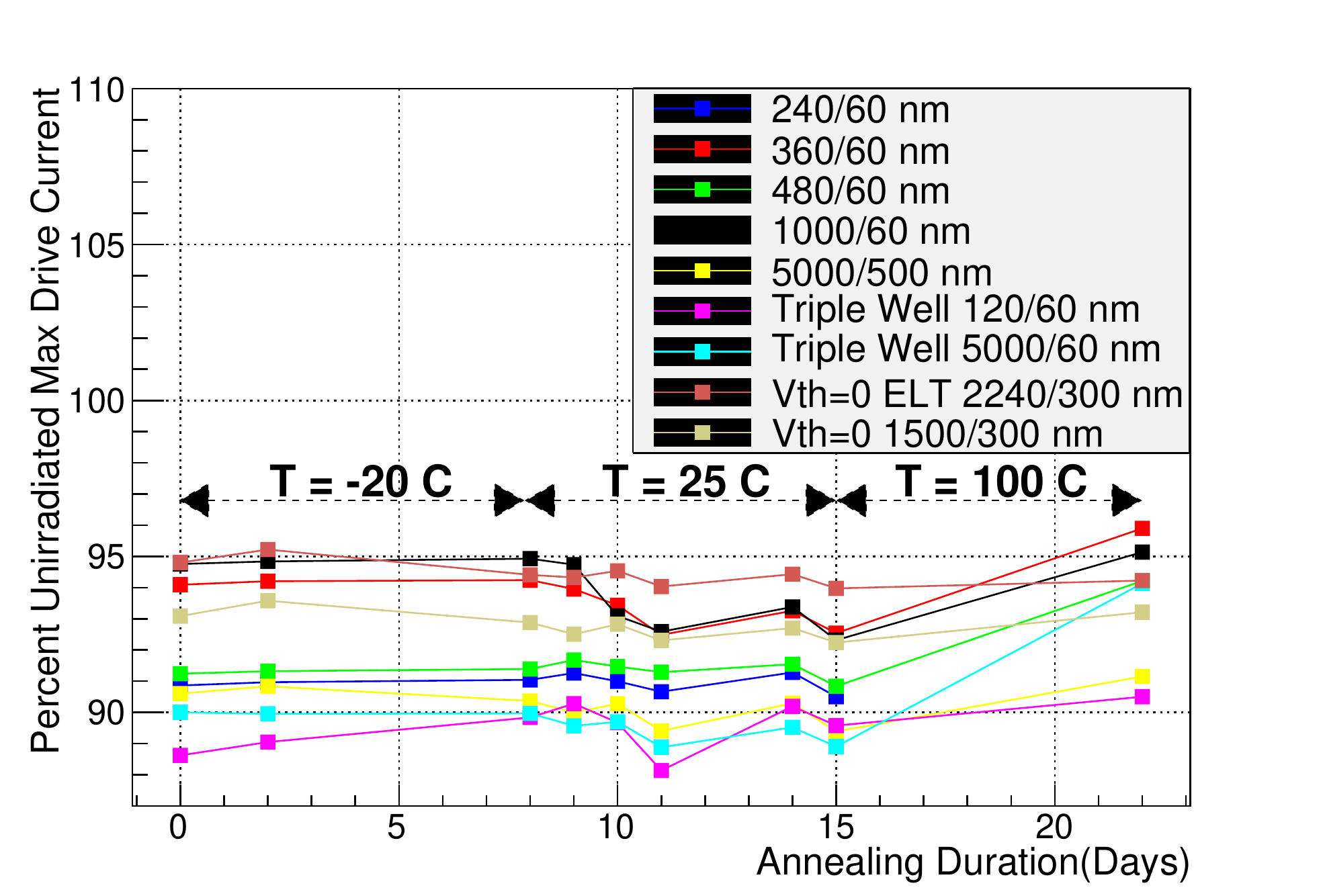}
\end{minipage}
\caption{The graph on the left shows the loss in maximum drain-source current after each irradiation step for 9 NMOS core transistors. The graph on the right shows the change in maximum drain-source current for the same 9 transistors during and after annealing.}
\label{fig:MaxCurDrive_NMOS}
\end{figure}

\begin{figure}
\begin{minipage}[b]{0.5\textwidth}
	\centering
	\includegraphics[width=\linewidth]{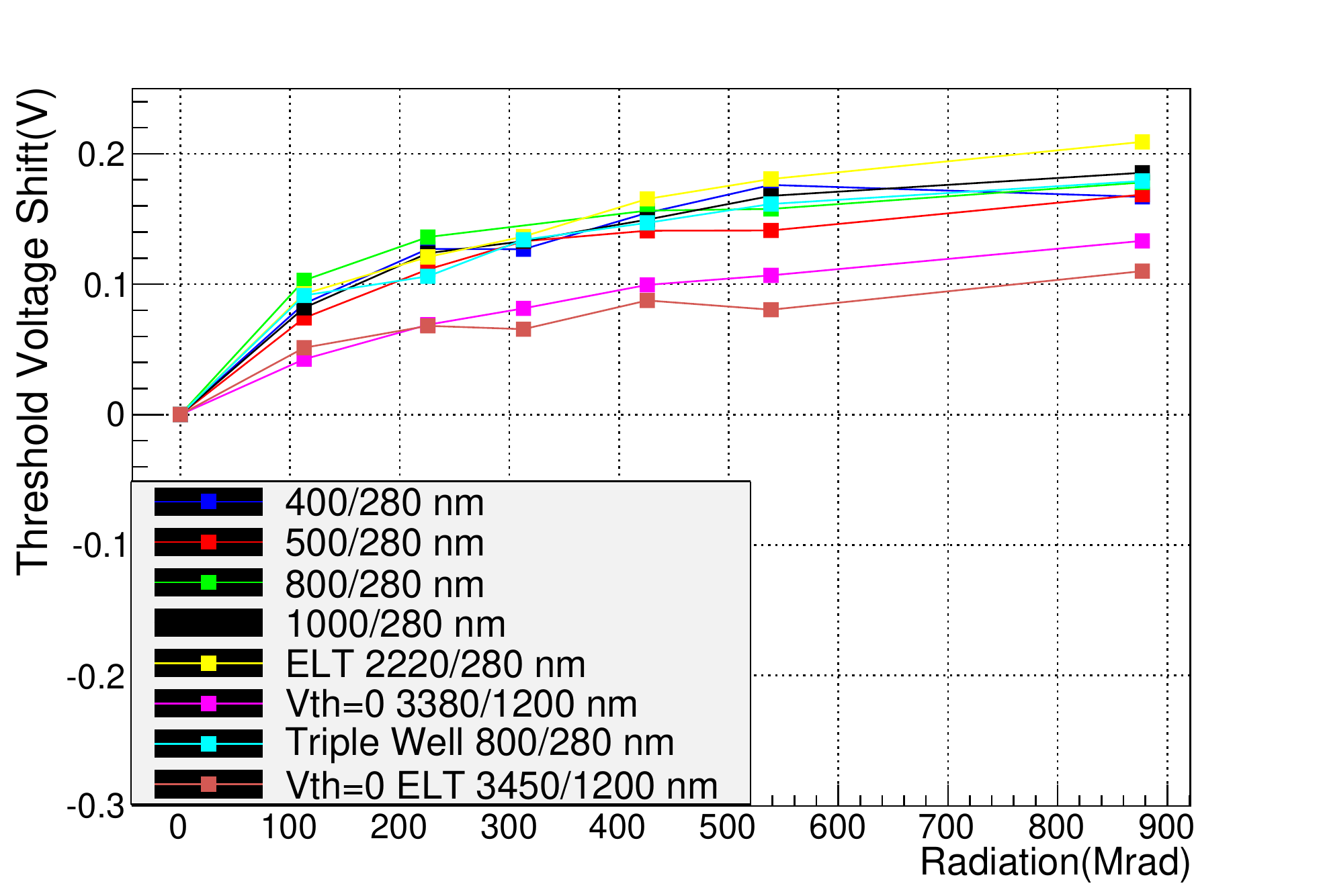}
\end{minipage}
\hspace{0.5cm}
\begin{minipage}[b]{0.5\textwidth}
	\centering
	\includegraphics[width=\linewidth]{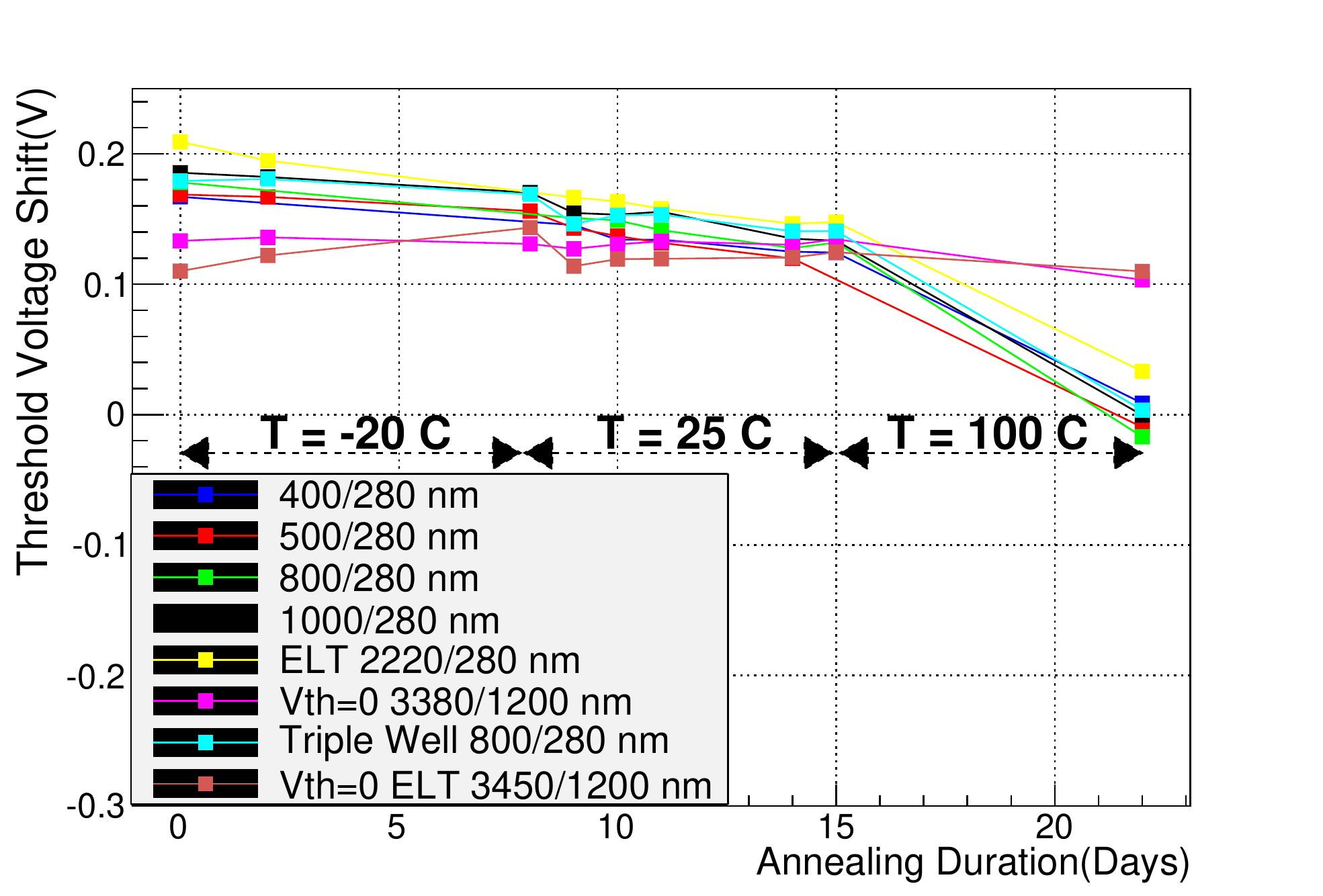}
\end{minipage}
\caption{The shift in threshold voltage for 8 NMOS I/O transistors irradiated to 878 MRad is shown in the graph on the left, while the graph on the right shows $V_{th}$ for the same 8 transistors during and after annealing.  No significant annealing was observed for the two zero $V_{th}$ I/O transistors.}
\label{fig:DGNMOS_Vth}
\end{figure}

Figure~\ref{fig:AnnealSuperpositionPlots} demonstrates the annealing effects observed in our data. Both the PMOS core transistors and the NMOS I/O transistors recovered significantly during the annealing period. 

Figures~\ref{fig:MaxCurDrive_PMOS} and~\ref{fig:MaxCurDrive_NMOS} show the evolution of the maximum drain-source current for a representative selection of PMOS and NMOS core transistors during irradiation and annealing.  We did not observe any significant differences in the effect of radiation on the various different types of NMOS transistors tested (normal layout, enclosed layout, triple well, and zero $V_{th}$).
Figure ~\ref{fig:DGNMOS_Vth} shows the threshold shift of a representative selection of NMOS I/O transistors during irradiation and annealing.

\section{Summary}

Previous measurements have established 65 nm CMOS as the leading candidate technology for HL-LHC electronics.
After an exposure of 200 Mrad, Bonacini, $\textit{et al.}$ reported \cite{Bonacini}, with one exception, only minor changes in transistor parameters.
The exception was a significant loss of maximum drain-source current by narrow PMOS core transistors.   They reported a 50\% reduction in maximum drive current for a 120/60 PMOS core transistor and a 35\% loss for a 360/60 PMOS core transistor.
This irradiation of ``cold'' 65 nm CMOS transistors was motivated by a concern that damage to pixel vertex detector readout electronics operated at $-20\,^{\circ}\mathrm{C}$ might be greater than observed in room temperature irradiations.  Our measurements show the same pattern of effects as observed previously, but the damage is less severe than was observed at room temperature, rather than more severe.

\section{Acknowledgments}

We wish to thank Charles Bowen of the University of Colorado
Department of Physics Precision Machine Shop, Nina Moibenko of Fermilab's Electrical Engineering Department,
and Donald Hanson, 
Maryla Wasiolek, and Nathan Hart of the Sandia National Laboratories Gamma
Irradiation Facility.
This work was supported in part by Department of Energy grant (DE-SC0006963).  Fermilab is operated by Fermi Research Alliance, LLC under Contract No. DE-AC02-07CH11359 with the United States Department of Energy.

\end{document}